\documentstyle[12pt]{article}
\begin{document}
\rightline{FTUV/93-37}
\rightline{IFIC/93-20}

\begin{center}
{\LARGE {\bf ON COHERENT STATES AND\\

\vspace{0.3cm}

 q-DEFORMED ALGEBRAS}}\footnote{ Presented at the 'International
Symposium on Coherent States' June 1993, USA.}

\end{center}

\vspace{0.5cm}

\begin{center}
{\Large{\bf  Demosthenes Ellinas }}\footnote{ Supported by DGICYT, Spain}
\end{center}

\vspace{0.3cm}

\begin{center}
Departamento de F\'{\i}sica Te\'orica and IFIC\\
Centro Mixto Universidad de Valencia-CSIC\\
E-46100 Burjassot (Valencia), Spain\\
ellinas@evalvx.ific.uv.es
\end{center}

\vspace{9cm}

\begin{abstract}
{\small{We review some aspects of the relation between ordinary coherent
states and q-deformed generalized coherent states with some  of the simplest
cases of quantum Lie algebras. In particular, new  properties of
(q-)coherent states are utilized to provide a path integral formalism
allowing to study a modified form of q-classical mechanics, to probe some
geometrical consequences of the q-deformation and finally to construct
Bargmann complex analytic realizations for some quantum algebras.}}

\end{abstract}

\newpage

\section{Introduction}

This work is devoted to the relation between q-continuous representations, or
q-generalized coherent states, and some of the simplest deformed Lie
algebras$^{1,2,3}$,
i.e the q-oscillator, $su_{q} (2)$ and the $su_{q} (1.1)$$^{4,5,6}$.
We start by giving
some basic notions of
Hopf algebras, the underlying theory of quantum groups, such as co-product,
non-co-commutativity, R-matrix, quantum group and algebra duality etc,
by using the prototype of $su_{q} (2)$ algebra (Chapt. 2). Then
q-deformed coherent states$^{5,7}$ are  introduced and their completeness
relations$^{8,9,10}$,
eigenvalue problem, minimum uncertainty and related properties are
discussed$^{11,12}$
(Chapt.3). Usual i.e
non-deformed coherent states$^{13,14}$ are also used to provide complex
analytic
realizations of these algebras and their co-products$^{15}$. These realizations
are given
in terms of a series in powers of ordinary derivative operators which act on
the Bargmann-Hilbert space of holomorphic functions endowed with the
ususal integration
measure (Chapt. 4).
Subsequently ordinary and q-deformed  coherent states corresponding to the
deformed oscillator and the quantum $su (2)$ and $su (1,1)$ algebras are
utilized to introduce a path integral formalism for these algebras.
In the semiclassical limit,
the resulting classical mechanics is studied by evaluating the symplectic
forms and the metrics, for arbitrary values of the
q-deformation
parameters; both metrics are found to be modified by the deformation.
Moreover, evaluation of the Riemann curvature scalar reveals a non-constant,
deformation induced curvature, in the two dimensional subspace of the
Hilbert space spanned by the coherent states vectors and attributing in this
way a
geometrical role to the q-deformation. When  non-deformed
coherent states are employed alternatively for the construction of the path
integrals,
the geometries remain unaffected by the deformation$^{12}$.The latter modifies
only
the upper symbol of each quantum algebra generator, which is given as
power series of the respective upper symbols of the non-deformed algebra
generators.
This result is further used in studying non-linear q-classical dynamics.
Especially
for this generalized q-mechanics the Liouville theorem of the
incompressibility of the phase space is valid (Chapt. 5).
Conclusions and future perspectives of the present work are finally
concentrated
in Chapter 6.

\section{Quantum Lie algebras}

Lef us start with the matrix

\begin{equation}
T = \left(
\begin{array}{cc}
a & b\\
c & d
\end{array} \right)
\end{equation}
the entries of which are generating elements of the associative  algebra of
functions of the quantum group $SU (2)$ denoted by $Fun(SU_{q} (2))$.
The six commutation relations, defining the algebra  $Fun(SU_{q}(2))$,
$(\omega = q - q^{-1})$,

\begin{equation}
ab=qba \; ac=qca \; bd=qdb \; cd=qdc \\
bc=cb \; ad-da={\omega }bc,
\end{equation}

\noindent
are written by using two auxiliary 4x4 matrices  $T_{1} = T \otimes 1 \,
,  \, T_{2} =
1 \otimes T$ and the R-matrix

\begin{equation}
R = \left(
\begin{array}{cccc}
q &  & & \\
& 1 & & \\
& \omega & 1 &\\
& & & q
\end{array}
\right) \, ,
\end{equation}

\noindent
in the compact form

\begin{equation}
R T_{1} T_{2} = T_{2} T_{1} R.
\end{equation}

The unitarity condition $T^{+} = T^{-1}$

\begin{equation}
\left( \begin{array}{cc}
\bar{a} & \bar{c}\\
\bar{b} & \bar{d}
\end{array} \right)
\,
=
\,
\left( \begin{array}{cc}
d & -q^{-1}b\\
-qc & a
\end{array} \right)
\end{equation}

\noindent
(where bar indicates an involution of the algebra of functions on the group
which changes the order of factors in a product), leads to the restrictions
$d = \bar a \, , \, b = - q \bar{c}$, where $q = e^{\gamma} \in {\cal R}$.
The unimodularity condition uses the quantum determinant and implies

\begin{equation}
Det_{q} T \equiv ad - qbc = a \bar{a} + q^{2} \bar{c} c = 1.
\end{equation}

Mathematically the $Fun(SU_{q}(2))$, is characterized as a Hopf algebra
(see details in Ref.(1-3); this means that three more operations can be
defined on the algebra. Indeed
if  $T = \{ t^{i}_j\}$ and the t-elements
are identified with the entries of the T-matrix in Eq.1, then
there exists the operation of  co-product $\Delta T = T \otimes T$, which in
components reads $\Delta (t^{i}_j)
= t^{i}_j \otimes t^{k}_j$,
then the
co-unit $\varepsilon (T) = 1$ which means that $\varepsilon (t^{i}_j) =
\delta^{i}_
j$ and finally, the co-inverse or antipode of the entries of T-matrix,
$S (T) = T^{-1}$, which has the
property $s (T) T = 1 = T s (T)$, but in general $S^{2} \not=1$.
Then the algebra
$Fun(SU_{q}(2)$, is said to be a non-commutative and non-co-commutative Hopf
algebra.
Moreover the R-matrix satisfies the Yang-Baxter equation $R_{12} R_{13} R_{23}
= R_{23}
R_{13} R_{12}$, where $R_{ij}$ is the R-matrix acting nontriavilly in the
ij-subspace of the triple tensor product.

The following property of the T-matrices can be qualified as the group property
of the quantum group $Fun(SU_{q}(2))$ and reflects the group property
of the underlying quantizing group: if the entries of $T$ and $T'$ are
commuting
and each of them satisfies Eq. (4) then their product $T'' = T T'$ also
satisfies
the same equation i.e,

\begin{equation}
R T'' _{1}T''_{2} = T''_{2} T''_{1} R.
\end{equation}

Finally the representation theory of the $Fun(SU_{q}(2))$ algebra$^{16,17}$
 states
that there two families of irreducible representations, both labelled
my a real index $0< {\phi}<{2 \pi}$. A trivial one $a=e^{i \phi}$, $c=0$ and
a infinite dimensional one in the Fock representation space of the usual
harmonic oscillator, $[A,{\bar A}]=1$, $N={\bar A}A$,  i.e

\begin{equation}
a=\sqrt {\frac {1-q^{2(N+1)} } {N+1}  }A  \; , \; c=e^{i \phi}q^{N}\,.
\end{equation}

The introduction of the quantum algebra $U_{q} (su (2))$  i.e, the deformed
enveloping algebra of $su (2)$ can be done via the construction of duality
pairs between  Hopf algebras . Defining, $R^{+} \equiv  R$ and $R^{-1} \equiv
P R^{-1} R$ where  P is the permutation matrix of two factors, or in
components $(R^{-})_{jl}^{ik} = (R^{-1})^{ki}_{ej}$ and forming the matrices,

\begin{equation}
L^{+} = \left(
\begin{array}{ll}
q^{J^{q}_{3}} & -\omega J^{q}_{+}\\
0 & q^{-J^{q}_{3}}
\end{array}  \right)
\;
,
\;
L^{-} = \left(
\begin{array}{ll}
q^{-J^{q}_{3}} & 0\\
- \omega J_{+}^{q} & q^{J_{3}^{q}}\\
\end{array} \right)
\end{equation}

\noindent
with entries in the $U_{q} (su(2))$ algebra, we set up the pairing between
quantum group and quantum algebra elements as $(L^{\pm } =\{ l^{\pm i}_{j} \})$

\begin{equation}
< T \otimes L^{\pm} > = R^{\pm} \;, \; < 1, L^{\pm} > = 1 \;,\; < T, 1> = 1 \,
,
\end{equation}

\noindent
 or in component form

\begin{equation}
<t^{i}_{j}, l^{\pm k}_{l} > = (R^{\pm})^{ik}_{jl} \,  \, <t^{i}_{j}, 1> =
\delta^{i}_{j} = < 1, l^{\pm i}_{j}>.
\end{equation}

Using further the properties of the pairing bracket, such as,
 $<t^{i}_{j} t^{k}_{l},
l^{\pm r}_{s} > = < t^{i}_{j} \otimes t^{k}_{l}, \Delta (l^{\pm r}_{s})>$
and $< t^{i}_{j}, l^{\pm k}_{l} l^{\pm r}_{s} > = < \Delta (t^{i}_{j}), l^{\pm
k}_{l} \otimes l^{\pm r}_{s} >$, the commutation relations of the algebra are
deduced $(\varepsilon = + + , - - , -+)$:

\begin{equation}
L_{1}^{\varepsilon} L_{2}^{\varepsilon} R = R L_{2}^{\varepsilon} L_{1}^
{\varepsilon},
\end{equation}

\noindent
where as before $L_{1}^{\varepsilon} = L^{\varepsilon} \otimes 1, L_{2} =
1 \otimes L^{\varepsilon}$. The so constructed quantum algebra
posseses the structure of a Hopf
algebra as well, with co-product $\Delta L^{\pm} =L^{\pm} \otimes L^{\pm}$, co-
unit $\varepsilon (L^{\pm}) = 1$ and antipode, $s(L^{\pm}) =
(L^{\pm})^{-1}$. From the unitarity of the quantum group matrix $T$ and
the duality connecting group and algebra, follows that $(L^{+})^{+}=
(L^{-1})^{-1}$,  which implies for the elements of the quantum $su(2)$ algebra
that
$\bar{J_{3}^{q}} = J_{3}^{q}$ and $\bar{J^{q}_{\pm}} = J^{q}_{\mp}$.

The explit form of the commutation relations (12) is,

\begin{equation}
[ J^{q}_{3}, J_{\pm}^{q}] = \pm J^{q}_{\pm} \quad , \quad
 [J_{+}^{q}, J_{-} ^{q}] =
[2 J_{3}^{q}]
\end{equation}

\noindent
where $[x] = \frac{q^{x} - q^{-x}}{q - q^{-1}} = \frac{ sinh \gamma x}{
sinh \gamma}$, or if $J_{1}^{a} = \frac{1}{2} (J_{+}^{q} + J_{-}^{q})$ and
$J_{2}^{q} = \frac{1}{2 i} (J_{+}^{q} - J_{-}^{q})$ we obtain the
commutation relations,

\begin{equation}
[J_{1}^{q}, J_{2}^{q}] = i [J_{3}^{q}], \quad  [J_{2}^{q} J_{3}^{q}] =
i J_{1}^{q},
\quad  [J_{3}^{q}, J_{1}^{q}] = i J_{2}^{q}\, .
\end{equation}

Writing explicitly the co-product of the algebra yields

\begin{equation}
\Delta J_{\pm}^{q} = J_{\pm}^{q} \otimes q^{J^{q}_{3}} + q^{-J_{3}^{q}}
\otimes J_{\pm}^{q}\,
\end{equation}

\begin{equation}
\Delta J_{3}^{q} = J_{3}^{q} \otimes 1 + 1 \otimes J_{3}^{q}\, .
\end{equation}

This co-product permits to define representation of the algebra ( Hopf algebra
)
in $V_1{\otimes}V_2$ tensor product of two representations $V_1$, $V_2$. The
irreducible representations of the $su_q(q)$ algebra generators are similar to
the usual ones,

\begin{equation}
J_{3}^{q} | jm > = m | jm>
\end{equation}

\begin{equation}
J_{\pm}^{q} | jm > = \sqrt{[j \pm m]\  [j \pm m + 1]} |j m \pm 1 >\, ,
\end{equation}

\noindent
from them we deduce the deforming mapping connecting deformed, indexed by $q$
and non-deformed algebra
generators ,

\begin{equation}
J_{+}^{q} = J_{+} \sqrt{\frac{[J_{3} + j + 1][J_{3} - j]}{(J_{3} + j + 1)
(J_{3} - j)}} \equiv J_{+} F (J_{3}), \quad  J_{-}^{q} = \bar{J_{+}^{q}},
\quad  J_{3}^{q}=J_{3} .
\end{equation}

The corresponding formulae for the $U_{q} (su (1,1))$ algebra
include the commutation relations,

\begin{equation}
[K_{3}^{q}, K_{\pm}^{q}] = \pm K_{\pm}^{q} \quad , \quad
[K_{+}^{q}, K_{-}^{q}] = - [2 K_{3}^{q}]
\end{equation}

\noindent
the co-products

\begin{equation}
\Delta K_{\pm}^{q} = K_{\pm}^{q} \otimes q^{K_{3}^{q}} + q^{-K_{3}^{q}}
\otimes K_{\pm}^{q} \quad , \quad \Delta K_{3} = K_{3} \otimes 1 + 1 \otimes
K_{3}
\end{equation}

\noindent
while the deforming mappings are,

\begin{equation}
K_{+}^{q} = K_{+}  \sqrt{\frac{[K_{3} - k + 1][K_{3} + k]}{(K_{3} - k + 1)
(K_{3}
+ k)}} \equiv K_{+} F (K_{3}) \quad , \quad K_{-}^{q} = \bar{K_{+}^{q}}.
\end{equation}

Finally, for the case of the q-oscillator the commutation relations are

\begin{equation}
a_{q} a_{q}^{+} - q a_{q}^{+} a_{q} = q^{-N} \quad , \quad [N, a_{q}^{+}] =
a_{q}^{+} \, , [N, a_{q}] = - a_{q}
\end{equation}

\noindent
while the deforming mappings deduced from the Fock representation read,

\begin{equation}
a_{q} = a \sqrt{\frac{[N]}{N}} \equiv a F(N) \quad , \quad a_{q}^{+} =
\sqrt{\frac{[N]}{N}} a^{+}.
\end{equation}

Let us remark finally that for the q-oscillator there is no  satisfactory
bialgebra and Hopf alegebra structure.

\section{Deformed coherent states and some of their properties}

 Deformed coherent states for the $q$-oscillator algebra are defined by
$(\alpha{\in {\cal C}})$

\begin{equation}
|\alpha)_q=e^{\alpha a^+_q}_q|0>=e^{\alpha T^+_a}|0>=\sum\limits^\infty
_{n=0}\frac{\alpha^n}{\sqrt{[n]!}}\ |n>=e^{\alpha T^+_a}e^{- \alpha a^+}
|\alpha)
\end{equation}

where

\begin{equation}
T^+_a=a^+_q\frac{(N+1)}{[N+1]}\hspace{1.5cm}{\rm and}\hspace{1.5cm}
T^-_a=\frac{(N+1)}{[N+1]}a_q,
\end{equation}

\noindent
and $[n]!=[1][2]...[n]$ ,
with the $q$-exponential function $e^x_q=\sum\limits^\infty_{n=0}
\frac{x^n}{[n]!}$. Let us remark that the introduction of the $q$-CS above
may involve deformed and non-deformed exponential functions$^{19,7}$ and that
the
latter possibility allows us to connect the undeformed oscillator CS to
the $q$-deformed one.
The normalized states
$|\alpha>_q=\frac{1}{\sqrt{_q\alpha|\alpha)_q}}|\alpha)_q$
with $_q(\alpha|\alpha)_q=e_q^{|\alpha|^2}$ are eigenstates of the annihilation
operator $a_q|\alpha>_q=\alpha|\alpha>_q$ and satisfy the completeness
relation

\begin{equation}
{\bf 1}=\int|\alpha)_q d\mu_q(\alpha)_q(\alpha|\hspace{1.5cm}{\rm with}
\hspace{1.5cm}d\mu_q(\alpha)=d^2_q{\alpha}e_q^{-|\alpha|^2}
\end{equation}

\noindent
where the integral is regarded as the Jackson $q$-integral.
Indeed if $\alpha=\sqrt{r}e^{\phi}$, the completeness integral is factorized
into an ordinary angular integral times a discrete Jackson integral, which
leads to the moment problem

\begin{equation}
\int_{0}^{\infty}e_q^{-r}r^{n}d_qr=[n]! \, .
\end{equation}

The derivation of this results in view of the fact that the $q$-exponential
function has zeros on the real axis and can even become negative, requires
some analysis which has been carried out in the literature.
Here will be sufficient to remind that the completeness condition requires the
antiderivative $D_x^{-1}$, of the discrete derivative,

\begin{equation}
D_x f(x)=x^{-1}[\frac{x\partial_x}{2}]f(x)=
\frac{f(q^{1/2}x)-f(q^{-1/2}x)}{(q^{1/2}-q^{-1/2})x},
\end{equation}

\noindent
which by virtue of formulae $[x,\partial_x]=1$ , $g(x\partial_x)x=
xg(x\partial_x +1)$ and $q^{k x\partial_x}f(x)=f(q^{k}x)$ , valid when
acting on analytic functions of variable x and by the expansion
$(2sinh(u/2))^{-1}=\sum_{l=0}^{\infty}e^{u(l+1/2)}$ , can be inverted as

\begin{eqnarray}
D_x^{-1}f(x) &=& x[\frac{x\partial_x+1}{2}]^{-1}f(x)=x(q^{-1/2}-q^{1/2})
                (2sinh\frac{-\gamma(x\partial_x+1)}{2})^{-1}f(x) \nonumber \\
             &=&
x(q^{-1/2}-q^{1/2})\sum_{l=0}^{\infty}q^{(l+1/2)}f(q^{(l+1/2)}x)\, .
\end{eqnarray}

These $q$-CS
are minimum-uncertainty states in the sence that they minimize the
$[q_q,p_q]$ commutator

\begin{equation}
\Delta q_q\Delta p_q=\frac{1}{2}|_q<\alpha|[q_q,p_q]|\alpha>_q|
\end{equation}

\noindent
where $a_q=\frac{1}{\sqrt{2}}(q_q+ip_q)$,  $a^+_q=\frac{1}{\sqrt{2}}
(q_q-ip_q)$.

A final comment before closing the oscillator case concerns the kind of
analytic functions as elements of the Bargmann -Hilbert space of
represantation  are
expandable in the $q$-CS basis. The square integrability condition of such
elements

\begin{equation}
< \psi | \psi> = || \psi ||^{2} = \frac{1}{\pi} \int d^{2}_{q} \alpha
(e^{-| \alpha|^{2}}_{q}) | \psi (\alpha) |^{2} < \infty
\end{equation}

\noindent
requires in the $q = 1$ case, that their growth  must
be of order $ 0 < \rho \leq 2$ and type $\tau = \frac{1}{2}$. However
due the fact that $e_{q}^{|\alpha|} \leq e^{|\alpha|}$, in the $q \not= 1$
case, the possible expandable functions  must be of slower growth
in comparison to those of the $q = 1$ case.
In light of some recent interest in the dynamics of
wavefunction zeros$^{18}$, it would be diserable to find the factorization
of these wavefunctions in terms of their roots.

 The deformed CS for the $su_q(2)$ algebra, related to representations
characterized by j=1/2, 1, 3/2,... are defined by
$(z{\in {\cal C}} )$

\begin{equation}
|z)=e^{zJ^+_q }_q|-j>=e^{zT^+_J}|-j>=\sum\limits^j_{m=-j}\left[\begin{array}
{c}2j\\j+m\end{array}\right]_q^{1/2}z^{j+m}|m>=e^{zT^+_z }e^{- zJ^+ }|z),
\end{equation}

\noindent
where the $q$-binomal is defined as $[^a_b]_q=\frac{[a]!}{[b]![a-b]!}.$
while

\begin{equation}
T^+_J=J^+_q\frac{(J^3_q+j+1)}{[J^3_q+j+1]}\hspace{1.5cm}{\rm and}
\hspace{1.5cm}T^-_J=\frac{(J^3_q+j+1)}{[J^3_q+j+1]}J^-_q.
\end{equation}

\noindent
The factor $(1+|z|^2)^{2j}_q\equiv\ _q(z|z)_q$ normalizes the states,
$|z>_q=\frac{1}{\sqrt{_q(z|z)_q}}|z)_q$ and using the general formula
\begin{equation}
(x+y)^n_q\equiv\sum\limits^n_{m=0}\left[\begin{array}{c}n\\m\end{array}
\right]_qx^{n-m}y^m=\prod\limits^n_{k=1}(x+q^{n-2k+1}y)
\end{equation}

\noindent
derived with the help of $[2m+1]=\sum\limits^m_{\ell=-m}q^{2\ell}$,
that factor is written as
\begin{equation}
_q(z|z)_q=\sum\limits^{2j}_{m=0}\left[\begin{array}{c}2j\\m\end{array}
\right]_q\ |z|^2m=\prod\limits^{2j}_{k=1}(1+q^{2j-2k+1}|z|^2)\ .
\end{equation}

\noindent
The normalized $q$-CS are complete with resolution of unity

\begin{equation}
{\bf 1}=\int|z)_qd\mu_q(z)_q(z|\hspace{1.5cm}{\rm with}\hspace{1.5cm}
d\mu_q(z)=\frac{[2j+1]}{_q(z|z)^2_q}d^2_qz\ ,
\end{equation}

\noindent
where again the Jackson $q$-integral is in use. This completeness
realation leads to a moment problem which entails the integral form
of the $q$-binomial,

\begin{equation}
\int_{0}^{\infty}r^n (1+r)^{-(2j+2)}d_qr=\frac{[n]![2j-n]!}{[2j+1]!}.
\end{equation}

\noindent
It is also interesting that the $q$-CS satisfy the eigenvalue problem,

\begin{equation}
(J^-_q+(q^j+q^{-j})z[J^3_q]-z^2J^+_q)|z>_q=0 \,
\end{equation}

\noindent
which upon taking the zero deformation limit reduces to its $q=1$
analogue equation which serves as a definition, up to a phase factor,
for the
$SU(2)$ coherent state$^{14}$. For future use we also record the formula

\begin{equation}
J^\pm_q|z>_q=z^{\mp 1}[j\pm J^3_q]|z>_q\ .
\end{equation}

 The coherent states related to the quantum $su(1,1)$ algebra
and associated with the discrete representations characterized by
k=1, 3/2, 2, 5/2, ...
are defined by the generators

\begin{equation}
T^+_K=K^+_q\frac{(K^3_q-k+1)}{[K^3_q-k+1]}\hspace{1.5cm}{\rm and}
\hspace{1.5cm}T^-_K=\frac{(K^3_q-k+1)}{[K^3_q-k+1]}K^-_q\ ,
\end{equation}

\noindent
in a manner similar to the previous cases:

\begin{equation}
|\xi)_q=e_q^{\xi K^+_q}|k;0>=e^{\xi T^+_K}|k;0>=\sum\limits^\infty
_{n=0}\frac{[2k+n+1]!}{[n]![2k+1]!}\xi^n|k;n>=e^{\xi T^+_K}e^{-\xi K^+}|\xi)\ ,
\end{equation}

\noindent
where $\xi\in D^k=\{|\xi|^2<q^{k-1}\}$.
With normalization factor obtained from the overlap of two states

\begin{equation}
(1-|\xi|^2)^{-2k}_q\equiv\ _q(\xi|\xi)_q=\sum\limits^\infty_{n=0}
\frac{[2k+n+1]!}{[n]![2k+1]!}|\xi|^{2n}
\end{equation}

\noindent
the normalized states are complete,

\begin{equation}
{\bf 1}=\int|\xi)_qd\mu_q(\xi)_q(\xi|\hspace{1.5cm}{\rm with}
\hspace{1.5cm}d\mu_q(\xi)=\frac{[2k-1]}{_q(\xi|\xi)^{-2}_q} d^2_q\xi\ ,
\end{equation}

\noindent
and obey the equations,
\begin{equation}
(K^-_q+(q^k+q^{-k})\xi[K^3_q]+\xi^2K^+_q)|\xi>_q=0
\end{equation}

\noindent
and

\begin{equation}
K^\pm_q|\xi>_q=\xi^{\mp 1}[K^3_q\mp k]|\xi>_q\ .
\end{equation}

\section{Bargmann realization for quantum algebras}

 The representation spaces of these deformed realizations
will be the ordinary Hilbert spaces of square-integrable analytic functions
$L^{2} ( \frac{G}{H}, d \mu (\zeta))$, built on the corresponding cosets
of the non-deformed Lie groups, i.e $G/H = \frac{Weyl-Heisenberg}{U(1)},
\frac{SU(2)}{U(1)}$ and $\frac{SU(1,1)}{U(1)}$. The invariant measure of
integration $d \mu (\zeta)$, is the so-called Bargmann measure. One feature of
the obtained
realizations of the quantum algebra generators is that they constitute a
deformation of the ordinary Lie algebra generators in the
sense
that they involve a series in powers of ordinary derivatives
(the coefficients of which depend on the deformation parameter
$q$) that reproduces in the  'classical'
$q\rightarrow 1$ limit the Lie algebra vector field
generators.
 Using generic symbols for the deformed algebra generators,
$G^{q}_{0} \equiv N, J_{3}, K_{3}$ and $G^{q}_{\pm} \equiv a^{\pm}_{q},
\; J^{q}_{\pm} \; ,
K^{q}_{\pm}$,  and for their co-products, which generically for our cases
can be written in the form

\begin{equation}
\Delta G^{q}_{\pm} = G^{q}_{\pm} \otimes g (G_{0}) + g' (G_{0}) \otimes
G^{q}_{\pm}
\end{equation}

\begin{equation}
\Delta G^{q}_{0} = G_{0} \otimes 1 + 1 \otimes G_{0} \quad ,
\end{equation}

\noindent
where $g(G_{0})$ and $g' (G_{0})$ are specific  for each algebra
separately.

As the co-product of each algebra generator acts in the tensor product
of the representation space of the algebra, we shall look for its
analytic functional realization carried by functions of two variables.
Ommiting the details of the construction, which can found in Ref.15,
we give only the final formulae, which for the realization of the step
generators of the algebras read

\begin{equation}
\pi_{\zeta} (G^{q}_{\pm}) = \tau^{\pm} \sum^{l}_{n=0} \frac{b^{\pm}_{n}}
{n!} \zeta^{n} \partial^{n}_{\zeta },
\end{equation}

\noindent
while for the co-products it is obtained that

\begin{equation}
\pi_{\zeta_{1} \zeta_{2}} (\Delta G^{q}_{\pm}) \Psi (\zeta_{1}, \zeta_{2})
= \left[ \pi_{\zeta_{1}} (G^{q}_{\pm}) \pi_{\zeta_{2}} (g (G_{0})) +
\pi_{\zeta_{1}}
(g' (G_{0})) \pi_{\zeta_{2}} (G^{q}_{\pm}) \right]
\Psi (\zeta_{1}, \zeta_{2}) \; ,
\end{equation}

\noindent
with

\begin{equation}
\pi_{\zeta_{1}} (g (G_{0})) = \sum^{l}_{m=0} \frac{c_{m}}{m!}
\zeta^{m}_{2} \partial^{m}_{\zeta_{2}} \quad ,
\end{equation}

\noindent
where the b's and c's are numerical factors, the limit $l$ has an appropriate
value
for each case and the $\tau$'s are specific for each algebra.

\section{Path integrals and q-deformed classical mechanics}

 We proceed now with the CS propagator utilizing the
coupleteness relations of the CS. Let $(\zeta=\alpha,z,\xi)$,
the transition amplitude between coherent states takes the form

\begin{equation}
{\bf K} = <\zeta''|U(t'',t')|\zeta'>=\int{\cal D}\mu(\zeta)\exp\biggl
[\sum\limits^{L}_
{\ell=1} \ell n <\zeta_\ell|\zeta_{\ell-1}>
-\frac{i}{\hbar}\varepsilon\frac{<\zeta_\ell
|H|\zeta_{\ell-1}>}{<\zeta_\ell|\zeta_{\ell-1}>}\biggr]
\end{equation}
where $\zeta_{0}\equiv{\zeta^{'}}$, $\zeta_{L}\equiv{\zeta^{''}}$,
$${\cal D}\mu(\zeta)=\lim\limits_{^{L\rightarrow\infty}_{\varepsilon
\rightarrow 0}}\prod\limits^{L-1}_{\ell=1}d\mu(\zeta_\ell)\ .$$

\noindent
and $\varepsilon=\frac{t''-t'}{L}$, while $H$ stands for a  Hamiltonian
 given in terms of generators of a quantum algebra.
In the continuous limit $\epsilon \rightarrow 0, L \rightarrow \infty$,
the lattice space path integral,
\begin{equation}
{\bf K} = \int{\cal D}\mu(\zeta) exp [
\frac{i}{\hbar} (\int_{0}^{t'' - t'} \theta - H (\zeta, \bar{\zeta}) dt)]
\end{equation}
where
$H (\zeta, \bar{\zeta}) = < H >$, and $ <(\cdot)> = < \zeta | (\cdot) |
\zeta >$, involves the usual one-form $\theta= i \hbar < \zeta | d | \zeta>$.
This one-form gives rise to a K\"{a}hler geometry i.e, the symplectic form and
the
metric of the generalized phase space, corresponding
to the cosets, $WH/ U(1) \approx R', \,   SU(2)/ U(1) \approx S^{2},
\,
 SU (1,1)/U(1) \approx S^{1,1}$ of the three groups in question$^{20}$.
 The effect
of deformation is only in the Hamiltonian which will involve the upper
symbols $< \zeta |$ gen $| \zeta>$, of  generators. These can be expressed
in terms of upper symbols of non-deformed generators, resulting in expressions
which manifest the nonlinear character of the deformation proccess. We take up
the $su_{q} (2)$ case and similar results hold for the two other cases too.
Indeed
for this case, employing the realizations of the previous section and the
formulae

\begin{equation}
z^{m} = \frac{< j - J_{3} >^{m}}{< J^{+}>^{m}} \quad , \quad < J_{\pm}^{m} > =
\frac{2j}{(2 j- m) ! 2 j^{m}} < J_{\pm}>^{m}
\end{equation}

\noindent
deduced by induction, we obtain

\begin{eqnarray}
< z | J^{q}_{\pm} | z> &=& \pi_{z} (J_{\pm}^{q}) (z|z) =
                            z^{\pm} \sum_{m = 0}^{2j}
\frac{b^{m}_{\pm}}{m!} z^{m} \partial _{z}^{m} (z|z) \nonumber \\
                       &=& \sum_{m = 0}^{2j} \frac{b_{m}^{\pm}}{2j^{m}}
       \left(^{2j }_{m} \right)
      < J_{-}>^{m} <j - J_{3} >^{m \pm 1} < J_{+} >^{-m \mp 1} \, .
\end{eqnarray}

Having used the ordinary CS in the path integral, the mechanics resulting
from the semiclassical form of the propagator is not modified by the
deformation. Indeed the $q$-deformation comes in only via the
modification of the classical Hamiltonian, which by the last formulae will
genarate an involved dynamics due to its non-linearity.

Alternatively, we can use the $q$-CS to bult up the path integral. The
lattice-space
form of the propagator is formally the same as it was priviously
but now the index $q$,
should be added everywhere to emphasize the use of $q$-CS. In the
continuous limit where $\epsilon \rightarrow 0, L \rightarrow \infty$,
while
the deformation parameter $q$ is retained fixed, assuming that
$\zeta_{\ell-1}\cong
\zeta_
\ell-\Delta \zeta_\ell$, then from the definition of the CS and the short-time
 approximation follows that

\begin{equation}
\varepsilon.\frac{1}{\varepsilon}\ell n (_q<\zeta_\ell|\zeta_{\ell-1}>_q)\cong
\frac{\varepsilon}{2}(\frac{\Delta\bar{\zeta}_\ell}{\varepsilon}
\  _q<\zeta_\ell|
T^-_i|\zeta_\ell>_q-\frac{\Delta \zeta_\ell}{\varepsilon}\
_q<\zeta_\ell|T^+_i|
\zeta_\ell>_q)\,
\end{equation}

\noindent
where $i=a,J,K$ and bar denotes complex conjugation. In the limit where
$L\rightarrow \infty $ and $\varepsilon\rightarrow 0$ the r.h.s of the
above expression is written formally as $\frac{1}{2}(\dot{\bar \zeta}
\ _q<\zeta|T^+_i|\zeta>_q
-\dot{\zeta} \ _q<\zeta|T^-_i|\zeta>_q)dt$, where overdot denotes time
derivative.
In that limit

\begin{equation}
{\bf K}=\int{\cal D}\mu_q(\zeta)\exp[\frac{i}{\hbar}\int\limits
^{t''-t'}_0\ dt\ L_q( \zeta, \bar{\zeta}; \dot{\zeta}, \dot{\bar{\zeta}}]
\end{equation}

\noindent
and the Langrangian is given by,

\begin{equation}
L_q=\frac{i\hbar}{2}[\dot{\zeta} \ _q<\zeta|T^+_i|\zeta>_q-\dot\zeta \ _q<\zeta
|T^-_i|\zeta>_q]
-{\cal H}_{q}
(\zeta,\bar{\zeta})\ ,
\end{equation}

\noindent
where ${\cal H}_{q}=\ _q<\zeta|H|\zeta>_q$, from which we extract the canonical
1-form
\begin{equation}
\theta_q=i\hbar \ _q<\zeta|d|\zeta>_q=
\frac{i\hbar}{2}\ (_q<\zeta|T^+_i|\zeta>_q d\zeta- \ _q<\zeta|T^-_i|\zeta> _q
d\bar{\zeta}) .
\end{equation}
By using the properties of the $q$-CS as above we obtain for the three
cases ,

\begin{equation}
L=\frac{i\hbar}{2} _q\biggl<\frac{N+1}{[N+1]}\biggr>_q
(\dot{\alpha}\bar{\alpha}
-\dot{\bar{\alpha}}\alpha  )-{\cal H}(\alpha,\bar{\alpha})\ ,
\end{equation}
and
\begin{equation}
L=\frac{i\hbar}{2}\ _q<J^3_q+j>_q(\dot{z}z^{-1}-\dot{\bar{z}}{\bar{z}^{-1}})
-{\cal H}
(z,\bar{z})\ ,
\end{equation}
and
\begin{equation}
L=\frac{i\hbar}{2}\ _q<K^3_q-k>_q(\dot{\xi}\xi^{-1}-
\dot{\bar{\xi}}{\bar{\xi}^{-1}}
 )-{\cal H}
(\xi,\bar{\xi})\ .
\end{equation}

The closed generalized symplectic 2-form $\omega_q=d\theta_q$, is obtained by
differentiation of the canonical 1-form above which we cast into the form,

\begin{equation}
\theta_q=\frac{i\hbar}{2}\frac{1}{_q(\zeta|\zeta)_q}
\frac{\partial_q(\zeta|\zeta)_q}{\partial|\zeta|^2}
(\bar{\zeta}d\zeta-
\zeta d\bar{\zeta}).
\end{equation}

\noindent
This yields the form,

\begin{eqnarray}
\omega_q &=& i\hbar(\ _q<T^-_iT^+_i>_q-\ _q<T^-_i>_q \ _q<T^+_i>_q)
\ d\bar{\zeta}\wedge d{\zeta} \nonumber \\
         &=& i\hbar{\frac{\partial}{\partial|\zeta|^2}}
[\frac{|\zeta|^2}{_q(\zeta|\zeta)_q}
\frac{\partial_q(\zeta|\zeta)_q}{\partial|\zeta|^2}]
\ d\bar{\zeta}\wedge d{\zeta} \, .
\end{eqnarray}

The modification of the resulting classical mechanics is obvious
in the last formula from the appearance of the deformed overlaps of CS, which
have
been given earlier. Along with the symplectic metric the metrical distance
gets modified.Proceeding by analogy with the case $q=1$, writing the potential
$\Phi_q=\ell n \ _q(\zeta|
\zeta)_q$,
the metric is written as
$ds^2=k \partial_{\zeta}\partial_{\bar{\zeta}}\Phi_q d{\zeta}d\bar{\zeta}$,
where $k$ is a proportionality factor. Using the definition of $q$-CS we find
that

\begin{eqnarray}
ds^2 &=& ( _q<T^-_iT^+_i>_q-\ _q<T^-_i>_q \ _q<T^+_i>_q)
        d\bar{\zeta} d{\zeta} \nonumber \\
     &=& i\hbar{\frac{\partial}{\partial|\zeta|^2}}
[\frac{|\zeta|^2}{_q(\zeta|\zeta)_q}
\frac{\partial_q(\zeta|\zeta)_q}{\partial|\zeta|^2}]
\ d\bar{\zeta} d{\zeta} \, .
\end{eqnarray}

The Riemann curvature scalar of that metric is not constant:

\begin{equation}
R=-(\partial_{\zeta}\partial_{\bar{\zeta}}\Phi_q)^{-1}{\ell n}
(\partial_{\zeta}\partial_{\bar{\zeta}}\Phi_q)\,;
\end{equation}

\noindent
for the $q$-oscillator for example, when the deformation parameter is small,

\noindent
$q=e^{\gamma}\approx1+\gamma$,

\begin{equation}
R=\gamma^212(1+2|\alpha|^2)\ .
\end{equation}

We conclude then that the $q$-deformation induces a non-constant curvature
in the two-dimensional space spanned by each $q$-CS vector, which is
a subspace of Hilbert space of representation for each algebra.The
symplectic metric, distance and the curvature scalar all refer to that subspace
labelling the $q$-CS, which in the non-deformed case coincides with the
respective coset spaces of each of the groups.
Here however the $q$-CS have been generated acting on a fiducial vector
with a displacement operator not originated from a coset decomposition
of some group element; with some abuse of language we call the labelling
space of our $q$-CS a deformed coset space, which as shown above is qualified
for generalized phase space.
The equation of motion in this phase space is derived by standard variation
of the Langrangian above in the limit when action $\ll{ \hbar}$, and reads,

\begin{equation}
i\hbar \dot{\bar{\zeta}}=(\ _q<T^-_iT^+_i>_q-\ _q<T^-_i>_q \ _q<T^+_i>_q)
^{-1}
\partial_{\zeta}{\cal H}.
\end{equation}

Finally, since we deal with generalized symplectic mechanics the Liouville
theorem holds true and we have verified explicitly that the phase space
area element, given by the 2-form above, is an invariant of the
canonical
equation of motion.

\section{Discussion}

The classical notion of coherent states is extented in the case of
quantum algebras in a rather fruitful way. These extensions offer
generalizations to known existing CS for Lie groups,
but also provide  $q$-CS as a tool for probing the novelties
of quantum algebras themselves. Based on the existing duality  between quantum
algebra and  its respective quantum group as Hopf algebras$^{21}$,
it would be possible
to construct on the co-representations of  quantum group, CS which
will be in duality to CS defined, as in this work, on representations
of the dual quantum
algebra. Unlike of course the CS of the algebra, the CS on its dual quantum
group will not have classical analogue.

Of particular importance is the use
of CS in the study of geometrical manifestations of the algebraic operation of
$q$-deformation, as it has been reviewed here. Based on the methodology of
coherent states and their direct relation to path integrals,
research can be carried out beyond the algebras studied here
to larger quantum algebras
and even to other generalized algebras associated with manifolds
with richer geometry and potential applications; such work will be
taken up elsewhere.


\newpage
\begin{thebibliography}{99}
\bibitem{1}
V.G.Drinfel'd,  Proc. ICM Berkeley CA (Providence, RI: AMS)
(A.M. Gleason Ed. 1986).
        .
\bibitem{2}
M.Jimbo,  Lett. Math. Phys. {\bf 10}(1985) 63; Comm. Math. Phys.
{\bf 102} (1986) 537.
\bibitem{3}
L.D.Faddeev N.Yu.Reshetikhin and L.A.Takhtadjan, Leningrad Math. J.
{\bf 1}(1990) 193.
\bibitem{4}
A.J.MacFarlane, J. Phys. {\bf A22} (1989) 4581.
\bibitem{5}
L.C.Biedenharn, J. Phys. {\bf A22} (1989) L871.
\bibitem{6}
M.Chaichian and P.P.Kulish, Phys. Lett. {\bf B234} (1990) 72.
\bibitem{7}
M.Chaichian, D.Ellinas and P.P.Kulish, Phys. Rev. Lett. {\bf 65} (1990) 980.
\bibitem{8}
R.W.Gray and C.A.Nelson, J. Phys. {\bf A23} (1990) L945.
\bibitem{9}
A.J.Bracken, D.S.McAnally, R.B.Zhang and  M.D.Gould, J. Phys. {\bf A24}
(1991) 1379.
\bibitem{10}
B.Jur\v{c}o, Lett. Math. Phys. {\bf 21} (1991) 51.
\bibitem{11}
D.Ellinas, Proc. of {\it XIX$^{th}$ ICGTMP},
June 1992 Salamanca, Spain.
Vol.I, p.99, ( M.A.del Olmo, M.Santander and J.Mateos Guilarte Eds., Anales
de Fisica, Monografias 1-2, CIEMAT,
Madrid 1993).
\bibitem{12}
D.Ellinas, J.Phys. {\bf A26} (1993) 553.
\bibitem{13}
J.R.Klauder,  {\it A coherent-states primer} in
 {\it Coherent states} (J.R. Klauder and B-S. Skagerstam Eds.,World Scientific
 Singapore 1985).
\bibitem{14}
A.M.Perelomov {\it Generalized coherent states
and their applications} (Springer Berlin, 1986).
\bibitem{15}
J.A.de Azca\'rraga and D.Ellinas, preprint FTUV/93-21, IFIC/93-09,
hep-th/9307083.
\bibitem{16}
S.Woronowicz, Comm. Math. Phys. {\bf 111} (1986) 613; Publ. RIMS Kyoto Univ.
{\bf 23} (1986) 117.
\bibitem{17}
L.L.Vaksman and Ya.S.Soibelman, Funct. Anal. Appl. {\bf 22} (1988) 170.
\bibitem{18}
D.Ellinas and V.Kovanis, preprint FTUV/93-13, hep-th/9309032,
Phys.Rev.A, in press.
\bibitem{19}
P.P.Kulish and E.V.Damaskinsky J. Phys. {\bf A23} (1990) L983.
\bibitem{20}
F.A.Berezin Comm. Math. Phys. {\bf 114} (1976) 153.
\bibitem{21}
S.Majid, Proc. of {\it XIX$^{th}$ ICGTMP},
June 1992 Salamanca, Spain.
Vol.I, p.145, ( M.A.del Olmo, M.Santander and J.Mateos Guilarte Eds., Anales
de Fisica, Monografias 1-2, CIEMAT,
Madrid 1993), and references quoted therein.

\end{thebibliography}
\end{document}